\documentclass[aps,preprint,showpacs,amsmath,amssymb]{revtex4-1}
\usepackage{graphicx}
\usepackage{dcolumn}% Align table columns on decimal point
\usepackage{bm}% bold math
\usepackage{hyperref}
\usepackage{xcolor}
\usepackage{soul}
\begin{document}

\title{Sudden change of genuine multipartite entanglement in non-Markovian dynamics }
\author{Mazhar Ali\footnote{Corresponding Author Email: mazharaliawan@yahoo.com or mazhar.ali@iu.edu.sa}}
\affiliation{Department of Electrical Engineering, Faculty of Engineering, Islamic University at Madinah, 
107 Madinah, Saudi Arabia}

%\date{\today}h

\begin{abstract}
We investigate entanglement dynamics of bipartite as well multipartite systems beyond Markov approximation. 
We study two pairs of cavity-reservoir systems, modeled as four qubits and track the change of 
entanglement among cavity-cavity qubits, reservoir-reservoir qubits, and also for genuine entanglement of 
all four qubits. 
For cavity-cavity qubits, we find that non-Markovianity prolongs the life of entanglement besides 
collapse/revival phenomenon. For reservoir-reservoir qubits, entanglement sudden birth is delayed 
accordingly along with oscillations. For all four qubits, with a specific initial state, we find that 
genuine entanglement develops gradually, reaches to a maximum constant value and then freezes at this value 
for some time before decaying. This sudden change in dynamics of genuine entanglement occurs for a time 
window where neither cavities nor reservoirs are entangled. In contrast to Markov process, sudden change 
phenomenon may be recurrent in non-Markovian regime.   
\end{abstract}


\maketitle

%%%%%%%%%%%%%%%%%%%%%%%%%%%%%%%%%%%%%%%
\section{Introduction}\label{S-intro}
%%%%%%%%%%%%%%%%%%%%%%%%%%%%%%%%%%%%%%%

Quantum entanglement is a precious resource for future technologies \cite{Erhard-NRP-2020,Friis-2019}. 
Several authors have devoted their time and energy to its characterization and detection 
\cite{Horodecki-RMP-2009,Guehne-PR474,Ali-QIP-2023,Zhang-AP-2023}.
However, this resource is fragile and may deplete from quantum states as a result of interaction with 
environment. There are many studies, which simulate the effects of various environments on entanglement of 
bipartite and multipartite systems  
\cite{Yu1,Yu2,Yu3,Yu4,Duer-PRL92,Hein-PRA71,Aolita-PRL100,Simon-PRA65,Borras-PRA79,Cavalcanti-PRL103,
Band-PRA72,Chaves-PRA82, Aolita-PRA82,Carvalho-PRL93,Lastra-PRA75,Guehne-PRA78,Lopez-PRL101,
Rau-EPL82,Ali-JPB42,Ali-JPB43,Ali-PRA81,Ali-JPB47,Weinstein-PRA85,Filippov-PRA88}.

For a specific scenario, the presence or absence of a photon in a cavity defines one qubit, 
whereas no photon or normalized collective state with single excitation in the reservoir defines the 
second qubit. Cavity-reservoir interaction leads to an entangled state. Two such systems effectively 
define a system of four qubits. We assume that there is no direct interaction between 
cavity qubits or reservoir qubits and there are no prior correlations among cavity-reservoir system. 
Entanglement dynamics of this system for various bipartitions have been 
investigated \cite{Lopez-PRL101-2008} for Markov process. It was found that whenever there is sudden 
death of entanglement among cavity qubits there is always sudden birth of entanglement among reservoir 
qubits. Another interesting possibility in this system is the presence of a time window where neither 
cavities nor reservoirs are entangled. The initial report of these findings did not explored dynamics 
of genuine entanglement but only bipartite entanglement among various bipartitions. Recently, we have 
studied dynamics of genuine entanglement and have found surprising results \cite{Ali-PRA90-2014}. 
For a specific initial state, genuine entanglement exhibits a sudden change in its dynamics 
and freezes onto a constant value for some time before decaying. Interestingly, such sudden change 
dynamics only happens during the time window where both cavity qubits and reservoir qubits are not 
entangled \cite{Ali-PRA90-2014}. Our findings were also verified and reported in 
experiments \cite{Aguilar-PRL113}. 

Recent progress on this problem investigate the extension of cavity-reservoir interaction beyond Markovian 
approximation. It was shown that similar to Markov process, there is sudden death of entanglement and 
corresponding sudden birth of entanglement \cite{Xu-PLA373-2009}. The authors studied dynamics of 
entanglement for several bipartitions of the systems. In this work, we extend our earlier investigations 
to non-Markovian process and study how entanglement evolves among cavity qubits, reservoir qubits and 
among four qubits in particular. We find that cavity-cavity entanglement is preserved for longer times as 
we move from Markov to non-Markov process. The entanglement also exhibits collapse/revival with decaying 
amplitude. For reservoir qubits, entanglement from cavity qubits is completely transferred in a smooth way 
for Markov process. For non-Markov process, entanglement show oscillations and gradually reaches its 
limiting value for small degree of non-Markovianity. For the case where there is a time window of no 
entanglement among cavities as well as reservoirs, we have surprising results. Among cavity qubits, we find 
that sudden death of entanglement is delayed as we move from Markov to non-Markov domain, however, there are 
no collapse and revival of entanglement. For reservoir qubits, we have sudden birth of entanglement in 
Markov and non-Markov domains and entanglement also show oscillations. The most interesting dynamics is
observed for genuine entanglement of four qubits. For a specific initial state, genuine entanglement 
exhibits sudden change dynamics for Markov as well as for non-Markov process. Interestingly, as we 
increase the degree of non-Markovianity, genuine entanglement show sudden change phenomenon two times and 
collapse/revival one time. We have not found sudden change dynamics for other quantum states. Recent 
progress in experiments studied this cavity-reservoir qubit beyond Markovian domain \cite{Silva-PRA101}. 
We believe that our finding can be easily verified in these experiments. 

This work is organized as follows. In section \ref{Sec:Model}, we briefly describe our model of interaction 
and briefly review the idea of genuine entanglement and its quantification. We present our main results in 
section \ref{Sec:results}. Finally, we conclude the work in section \ref{Sec:conc}. 

%%%%%%%%%%%%%%%%%%%%%%%%%%%%%%%%%%%%%%%%%%%%%%%%%
\section{Non-Markovian model of two qubits coupled to uncorrelated reservoirs} \label{Sec:Model}
%%%%%%%%%%%%%%%%%%%%%%%%%%%%%%%%%%%%%%%%%%%%%%%%%

We define two qubits system as follows. The cavity modes with absence or presence of a single photon is 
modeled as one qubit. Each mode interacts independently with its own reservoir, which can also be 
modeled as a second qubit described below. We have two such pairs of cavity-reservoir qubits in 
the system. The Hamiltonian responsible for the interaction between a single cavity mode and a $N$-mode 
reservoir is given by \cite{Lopez-PRL101-2008, Ali-PRA90-2014}
\begin{eqnarray}
\hat{H} = \hbar \omega \hat{a}^\dagger \hat{a} + \hbar \sum_{k = 1}^N \omega_k \hat{b}^\dagger \hat{b} 
+ \hbar \sum_{k = 1}^N g_k \big( \hat{a} \hat{b}_k^\dagger + \hat{b}_k \hat{a}^\dagger \big) \,,
\label{Eq:Hamil}
\end{eqnarray}
where the first two terms describe the single cavity mode with frequency $\omega$ and $N$-mode reservoir 
with frequency $\omega_k$, respectively. The last term is related with the interaction between cavity and 
$N$ modes of the reservoir. The Hamiltonian in the interaction picture with 
$\omega - \omega_k = 0$, can be written as
\begin{eqnarray}
\hat{H}_I = \hbar \sum_k g_k (\hat{a} \, \hat{b}_k^\dagger + \hat{a}^\dagger \, \hat{b}_k) \, .
\end{eqnarray}
This interaction picture Hamiltonian is the starting point of many studies in quantum optics and quantum information. This Hamiltonian is 
time-independent and will be used below to evaluate the time evolution of state vector. Another possibility is to introduce Heisenberg picture, where time-dependence is in the Hamiltonian and we can take the state vector as stationary. We find it simpler to consider the approach discussed below.

We consider the initial condition with cavity mode contains a single photon and its corresponding reservoir 
is in the vacuum mode, i.e., $|\psi(0)\rangle = |1 \rangle_C \otimes |\bar{{\bf 0}}\rangle_R $, 
where $ |\bar{{\bf 0}}\rangle_R = \Pi_{k=1}^N |{\bf 0}_k\rangle_R$ is the collective vacuum state of 
$N$-modes of reservoir. The time evolution of a state vector follows the Schr\"odinger equation 
\begin{equation}
 i \, \hbar \frac{\partial |\psi(t)\rangle}{\partial t	} = \hat{H}_I \,|\psi(t)\rangle\,. \label{Eq:SWE}
\end{equation}
At any time, we have the state vector  
\begin{eqnarray}
|\psi(t)\rangle = C_0(t)|1\rangle_C\,|\bar{{\bf 0}}\rangle_R +\sum_{k=1}^N C_k(t)\,|0\rangle_C\,
|{\bf 1}_k\rangle_R \,,
\label{Eq:TE}
\end{eqnarray}
where $|{\bf 1}_k\rangle_R$ describes the presence of a single photon in mode $k$, and $C_k(t)$ are 
the corresponding probability amplitudes. Substituting Eq.(\ref{Eq:TE}) into Eq.(\ref{Eq:SWE}) with the 
initial condition, we can write 
\begin{equation}
\dot{C}_0(t) = - \int_0^t dt' F(t-t') C_0(t') \, , 
\end{equation}
where the correlation function $F(t-t')$ in the limit of $N \to \infty$ is 
\begin{equation}
 F(t-t') = \int d\omega J(\omega) e^{i(\omega -\omega_0)(t-t')} \,,
\end{equation}
where $J(\omega)$ as spectral density of the reservoir. We consider the Lorentzian spectral distribution 
defined as 
\begin{equation}
 J(\omega) = \frac{\gamma_0}{2 \pi} \, \frac{\gamma^2}{(\omega - \omega_0)^2 + \gamma^2} \,,
\end{equation}
where $\gamma_0^{-1}$ and $\gamma^{-1}$ are qubit relaxation time and reservoir correlation time, 
respectively \cite{Breuer-book}.   
In non-Markovian regime, $\gamma_0 > \gamma/2$, whereas the condition $\gamma_0 < \gamma/2$ describes the 
Markovian process. The probability amplitude for this situation has been solved \cite{Xu-PLA373-2009}, and 
can be written as
\begin{equation}
C_0(t) = e^{- x \, \frac{\gamma_0 t}{2}} \, \bigg[ \, \cos\bigg(\frac{\sqrt{x(2-x)} \, \gamma_0 t}{2}\bigg) 
+ \frac{x}{\sqrt{x (2-x)}} \, \sin\bigg(\frac{\sqrt{x(2-x)} \, \gamma_0 t}{2}\bigg) \, \bigg] \, ,
\label{Eq:PA}
\end{equation}
where $x = \gamma/\gamma_0$ and non-Markovian regime restricts the parameter $x < 2$. With this result, 
the general solution in Eq.(\ref{Eq:TE}) can be written as 
\begin{eqnarray}
|\psi(t)\rangle = C_0(t) \, |1\rangle_C \, |\bar{{\bf 0}}\rangle_R 
+ C(t) \, |0\rangle_C \, |\bar{{\bf 1}} \rangle_R \, ,
\label{Eq:TEES}
\end{eqnarray}
where $|\bar{{\bf 1}} \rangle_R = 1/{C(t)} \, \big[ \, \sum_{k=1}^N C_k(t) \, |{\bf 1}_k\rangle \big]$ for 
$N \to \infty$. The probability amplitude $C(t) = \sqrt{1- C_0(t)^2}$ and Eq.(\ref{Eq:TEES}) describes an 
effective two-qubit system. We can take such pairs to define four qubits system ($2$ pairs), or six 
qubits system ($3$ pairs) and so on. In this study, we restrict ourselves to four qubits systems. 

We are now in a position to study entanglement dynamics of an arbitrary initial state of cavity-cavity 
qubits, the corresponding appearance of entanglement among reservoirs qubits, and behavior of genuine 
multipartite entanglement of respective collective state of four qubits. We take both reservoirs to be 
in vacuum state and without losing generality, we take two qubits in an arbitrary 
X-state \cite{ARPR09}. The initial state of four qubits, is given as  
\begin{eqnarray}
\rho_{tot} (0) = \varrho_X \otimes |\bar{0}_{r_1} \bar{0}_{r_2}\rangle\langle\bar{0}_{r_1} \bar{0}_{r_2}| \, , 
\label{Eq:XIS}
\end{eqnarray}
where  
\begin{eqnarray}
\varrho_X = \left( 
\begin{array}{cccc}
\rho_{11} & 0 & 0 & \rho_{14} \\ 
0 & \rho_{22} & \rho_{23} & 0 \\ 
0 & \rho_{32} & \rho_{33} & 0 \\
\rho_{41} & 0 & 0 & \rho_{44}
\end{array}
\right).
\label{Eq:Xs}
\end{eqnarray}
The unit trace and positivity conditions state that $\sum_{i=1}^4 \rho_{ii} = 1$,  
$ \rho_{22} \rho_{33} \geq |\rho_{23}|^2$, and $ \rho_{11} \rho_{44} \geq |\rho_{14}|^2$. 
$\varrho_X$ is entangled if and only if either 
$\rho_{22} \rho_{33} < |\rho_{14}|^2$ or $\rho_{11} \rho_{44} < |\rho_{23}|^2$. The orthonormal photonic 
eigenstates $|1\rangle = |0_A \, 0_B\rangle$, $|2\rangle = |0_A \, 1_B\rangle$,
$|3\rangle = |1_A \, 0_B \rangle$, $|4\rangle = |1_A \, 1_B\rangle$ form the basis for two qubits. 
It is simple to obtain the time evolved density matrices for cavity-cavity (CC) qubits and 
reservoir-reservoir (RR) qubits by taking respective partial traces. The cavity-cavity qubits 
density matrix is given as 
\begin{widetext}
\begin{eqnarray}
\rho_{C C}(t) = \left( 
\begin{array}{cccc}
\rho_{11}(t) & 0 & 0 & \rho_{14} \, C_0^2(t) \\ 
0 & (\rho_{22} + \rho_{44} \, C^2(t) ) \, C_0^2(t) & \rho_{23} \, C_0^2(t) & 0 \\ 
0 & \rho_{32} \, C_0^2(t) & (\rho_{33} + \rho_{44} \, C^2(t)) \, C_0^2(t) & 0 \\
\rho_{41} \, C_0^2(t) & 0 & 0 & \rho_{44} \, C_0^4(t)
\end{array}
\right),
\label{Eq:TECXs}
\end{eqnarray}
\end{widetext}
where $\rho_{11}(t) = \rho_{11} + ( \rho_{22} + \rho_{33} + \rho_{44} \, C^2(t) ) \, C^2(t)$. Similarly, 
reservoir-reservoir density matrix is given as 
\begin{widetext}
\begin{eqnarray}
\rho_{R R}(t) = \left( 
\begin{array}{cccc}
\sigma_{11}(t) & 0 & 0 & \rho_{14} \, C^2(t) \\ 
0 & (\rho_{22} + \rho_{44} \, C_0^2(t) ) \, C^2(t) & \rho_{23} \, C^2(t) & 0 \\ 
0 & \rho_{32} \, C^2(t) & (\rho_{33} + \rho_{44} \, C_0^2(t)) \, C^2(t) & 0 \\
\rho_{41} \, C^2(t) & 0 & 0 & \rho_{44} \, C^4(t)
\end{array}
\right),
\label{Eq:TERXs}
\end{eqnarray}
\end{widetext}
where $\sigma_{11}(t) = \rho_{11} + ( \rho_{22} + \rho_{33} + \rho_{44} \, C_0^2(t) ) \, C_0^2(t)$.
The time evolved combined state of four qubits can be easily obtained after some algebra. 

For two qubits (cavity-cavity or reservoir-reservoir), detection of entanglement can be done by various 
equivalent methods, like partial transpose moments 
\cite{Vidal-PRA65,Tran-2015,Enk-2012,Elben-PRA-2019,Ketterer-2019,Knips-2020,Gray-2018,Zhou-2020}, 
principal minors \cite{Ali-QIP-2023,Zhang-AP-2023,Huang-PRA76}, or concurrence \cite{Wootters-2001}. 
Eq.(\ref{Eq:PA}) poses one difficulty in obtaining the analytical expressions for any measure of 
entanglement because the eigenvalues of partially transposed matrix are not easy to obtain. However, we 
can easily study entanglement dynamics numerically using programs developed in MATLAB. 
In bipartite systems, any given quantum state is either entangled or separable. For multipartite systems, 
any given quantum state can be fully separable, bi-separable or genuinely entangled. Having negative partial 
transpose w.r.t any bipartition does not mean that quantum state is genuine entangled because there are 
quantum states which have negative partial transpose with respect to all bipartitions, nevertheless they 
are not genuinely entangled \cite{Horodecki-RMP-2009}. 

Let us clarify the ideas by taking three qubits case with obvious extension to $N$ parties and higher 
dimensions. A three-qubit state is said to be fully separable ($fs$) if 
$\varrho^{fs} = \sum_i p_i \varrho^A \otimes \varrho^B \otimes \varrho^C$. A pure state is 
bi-separable ($bs$) if $|\psi^{bs}\rangle = |\phi^A\rangle |\phi^{BC}\rangle$, where $|\phi^{BC}\rangle$ 
may be an entangled state for $BC$. A mixed state is bi-separable if 
$\varrho^{bs} = \sum_j p_j |\phi_j^{bs}\rangle\langle \phi_j^{bs}|$ 
for all three bipartitions. A state is said to be genuine entangled if it is not fully separable or 
bi-separable \cite{Bastian-PRL106-2011}. 
Detection of genuine entanglement \cite{Bastian-PRL106-2011} can be based on using positive partial 
transpose (PPT) mixtures. PPT mixtures have been characterized by semidefinite programming (SDP), however, 
it should be noted that there are genuine entangled states which are PPT mixtures \cite{Bastian-PRL106-2011}. 
Therefore, if the measure is positive for a given multipartite state then it is guaranteed to be genuine 
entangled, otherwise we should use some other techniques to check the entanglement properties wherever 
possible. It was shown \cite{Bastian-PRL106-2011} that a state is a PPT mixture iff the following 
optimization problem 
\begin{eqnarray}
\min {\rm Tr} (\mathcal{W} \varrho)
\end{eqnarray}
with constraints that for all bipartition $M|\bar{M}$
\begin{eqnarray}
\mathcal{W} = P_M + Q_M^{T_M},
 \quad \mbox{ with }
 0 \leq P_M\,\leq 1 \mbox{ and }
 0 \leq  Q_M  \leq 1\, 
\end{eqnarray}
has a positive solution. The constraints mean that the operator $\mathcal{W}$ is a decomposable entanglement 
witness for any bipartition. If this minimum is negative, then $\varrho$ is certified to be 
genuinely entangled. This minimum value is also an entanglement monotone \cite{Bastian-PRL106-2011}. We 
denote this measure by $E(\rho)$. For bipartite systems, this monotone is equivalent 
to {\it negativity} \cite{Vidal-PRA65}, whereas for multipartite systems, it can be called 
genuine negativity. It has advantage over all other measures that simple programs can be written to 
compute it numerically using MATLAB \cite{Bastian-PRL106-2011}.

We first provide an example of three qubits state which is PPT w.r.t. all three bipartitions but can be entangled 
(not detected by PPT-mixture). The state is given as \cite{Kay-PRA83-2011} 
\begin{eqnarray}
\varrho_{K} = \frac{1}{8+8\alpha} \left( 
\begin{array}{cccccccc}
4+\alpha & 0 & 0 & 0 & 0 & 0 & 0 & 2 \\ 
0 & \alpha & 0 & 0 & 0 & 0 & 2 & 0 \\ 
0 & 0 & \alpha & 0 & 0 & -2 & 0 & 0 \\
0 & 0 & 0 & \alpha & 2 & 0 & 0 & 0 \\
0 & 0 & 0 & 2 & \alpha & 0 & 0 & 0 \\
0 & 0 & -2 & 0 & 0 & \alpha & 0 & 0 \\
0 & 2 & 0 & 0 & 0 & 0 & \alpha &  0 \\
2 & 0 & 0 & 0 & 0 & 0 & 0 & 4+\alpha  \\
\end{array}
\right).
\label{Eq:Ak}
\end{eqnarray}
This matrix is valid quantum state for $\alpha \geq 2$ and it is fully separable for $\alpha > 2 \sqrt{2}$. It was 
shown numerically and analytically \cite{Guehne-PLA-2011} that the state is entangled for $2 \leq \alpha \leq 2 \sqrt{2}$. 

Another example of mixed state is given as \cite{Guehne-PR474}
\begin{eqnarray}
\rho = \frac{1}{3} (|\phi^+ \rangle \langle \phi^+ |_{AB} \otimes |0\rangle \langle 0|_C + 
|\phi^+ \rangle \langle \phi^+ |_{AC} \otimes |0\rangle \langle 0|_B + |\phi^+ \rangle \langle \phi^+ |_{BC} \otimes |0\rangle \langle 0|_A ) \, ,
\end{eqnarray} 
where $|\phi^+ \rangle = (|00\rangle + |11 \rangle)/\sqrt{2}$.
It is obvious from the construction of this mixed state that it is a mixture of biseparable states. A recent study analyzed the various classes of quantum states for tri-partite systems \cite{Schneeloch-PR2020}. Interestingly, the above mixed state initially thought to be biseparable, is so called 'fully inseparable' (For details, please see the Figure 1 in Ref. \cite{Schneeloch-PR2020}. It is easy to verify that all three partial transposes of this state w.r.t to each partition is negative, nevertheless the state is not genuinely entangled. 
%%%%%%%%%%%%%%%%%%%%%%%%%%%%%%%%%%%%%
\section{Results} \label{Sec:results}
%%%%%%%%%%%%%%%%%%%%%%%%%%%%%%%%%%%%%

As a first example, we consider cavity-cavity qubits to be in the pure state
\begin{eqnarray}
|\psi(0)\rangle =  \, \alpha \, |0 \, 0 \rangle \, + \, \beta \, |1 \, 1\rangle \, , 
\label{Eq:JIS}
\end{eqnarray}
with $|\alpha|^2 + |\beta|^2 = 1$. It is known that under Markovian dynamics, cavity qubits loose their 
entanglement (``entanglement sudden death") in a finite time for $\alpha < \beta$ 
\cite{Yu1,Yu2,Yu3,Yu4}. The entanglement transferred to reservoir qubits may appear at a time 
(``entanglement sudden birth") depending on the relation between $\alpha$ and $\beta$. These two times 
are equal for $\beta = 2 \, \alpha$. The reservoirs may get entangled before the cavities disentangle 
for $\beta < 2 \, \alpha$ or the reservoirs' entanglement may come after the cavities have disentangled 
for $\beta > 2 \, \alpha$. This last possibility has the peculiarity of a time window where both cavity 
qubits and reservoir qubits are disentangled \cite{Lopez-PRL101-2008}. 

\begin{figure}[h]
\centering
\scalebox{2.28}{\includegraphics[width=1.99in]{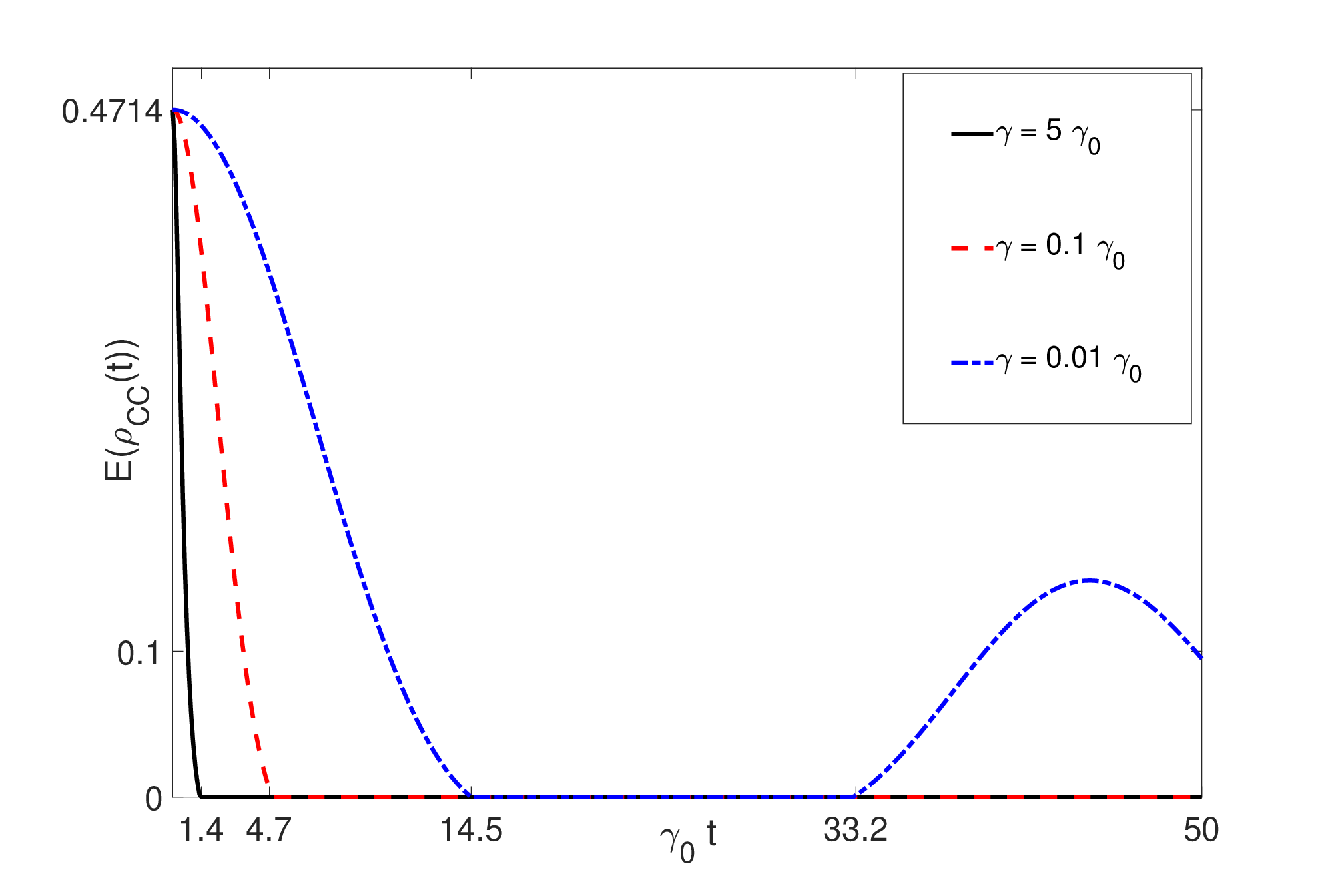}}
\caption{(Color online) $E$-monotone is plotted against parameter $\gamma_0 t$ for cavity-cavity qubits. 
The black solid line is for Markovian dynamics. The red dashed line and blue dashed-dotted line both present 
non-Markovian dynamics. We have taken $\alpha = \sqrt{1/3}$ and $\beta = \sqrt{2/3}$. See text 
for more description.}
\label{FIG:CR1}
\end{figure}
In Figure (\ref{FIG:CR1}), we plot cavity-cavity entanglement for initial states in Eq.~(\ref{Eq:JIS}) 
against dimensionless parameter $\gamma_0 \, t$. We have taken $\alpha = \sqrt{1/3}$ and 
$\beta = \sqrt{2/3}$.  The solid line denotes Markovian dynamics (with $x = 5$). We can see that 
entanglement vanishes at a finite time $\gamma_0 \, t \approx 1.4$. In our earlier study of purely 
Markovian dynamics \cite{Ali-PRA90-2014}, this finite end time of entanglement for same initial state was 
little different but near to this point. This slight change in dynamics is due to wide range of Markovian 
process for $x > 2$. The dashed (red) curve line is for non-Markovian case with $x = 0.1$. We observe that 
life time of entanglement is prolonged up to $\gamma_0 \, t \approx 4.7$. It is clear that entanglement in 
this case does not reappear among cavity qubits for very long time. We are not sure whether entanglement 
may appear at later times.  
The dashed-dotted (blue) curve denotes entanglement for non-Markovian process with $x = 0.01$. We observe 
that life time of entanglement is further delayed up to $\gamma_0 \, t \approx 14.5$. However, in this 
situation entanglement reappears among cavity-cavity qubits at $\gamma_0 \, t \approx 33.2$, 
grows to some value with much lower peak and then again tends to vanish. We expect such collapse/revival 
of entanglement with diminishing peaks for other smaller values of $x$. 

\begin{figure}[h]
\centering
\scalebox{2.30}{\includegraphics[width=1.99in]{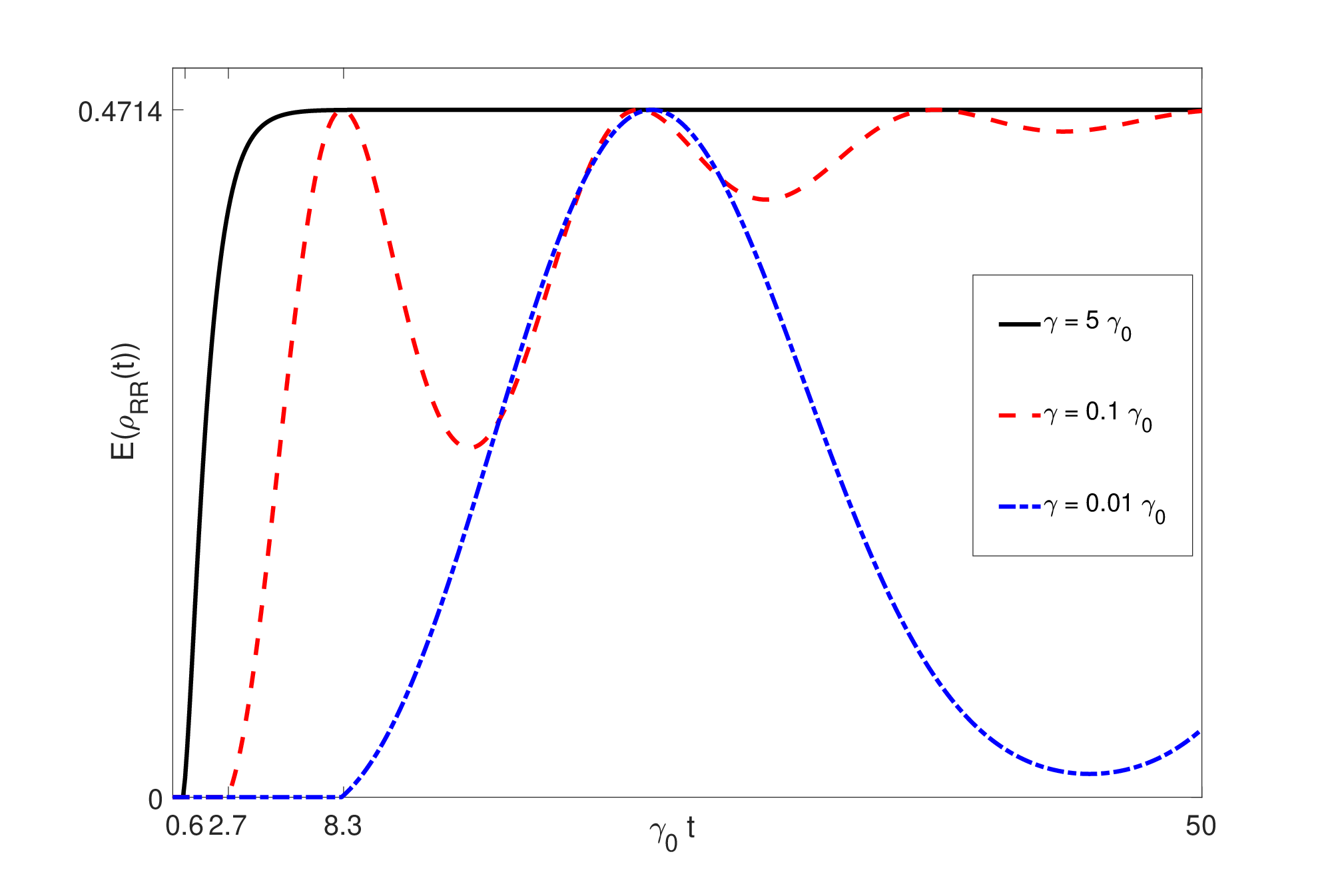}}
\caption{(Color online) Entanglement is plotted against parameter $\gamma_0 \, t$ for reservoir-reservoir 
qubits with $\alpha = \sqrt{1/3}$ and $\beta = \sqrt{2/3}$. 
The (black) solid line is for Markovian process whereas other two curves are for non-Markovian dynamics.}
\label{FIG:CR2}
\end{figure}
In Figure (\ref{FIG:CR2}), we plot reservoir-reservoir entanglement as it develops against parameter 
$\gamma_0 \, t$. We have started with initial condition $\alpha = \sqrt{1/3}$ and $\beta = \sqrt{2/3}$. 
As there is finite end of entanglement among cavity qubits (Figure \ref{FIG:CR1}), there is 
corresponding sudden birth of entanglement among reservoirs \cite{Lopez-PRL101-2008}.
The solid curve denotes Markov process with $x = 5$. We can see that as expected entanglement appears 
at $\gamma_0 \, t \approx 0.6$ and reaches to the maximum value of $0.4714$ and maintains this value till 
infinity. Hence all entanglement of cavity qubits is transferred to reservoirs in a finite time. 
The dashed (red) curve denotes non-Markovian process with $x = 0.1$. There are two observations for this 
case. First, sudden birth of entanglement is delayed to $\gamma_0 \, t \approx 2.7$. Second, there are 
oscillations of entanglement terminating at the maximum value. For an another instance of non-Markovianity 
with $x = 0.01$, dashed-dotted curve shows that entanglement appears at $\gamma_0 \, t \approx 8.3$ and 
oscillations have larger durations. 

\begin{figure}[h]
\centering
\scalebox{2.30}{\includegraphics[width=1.99in]{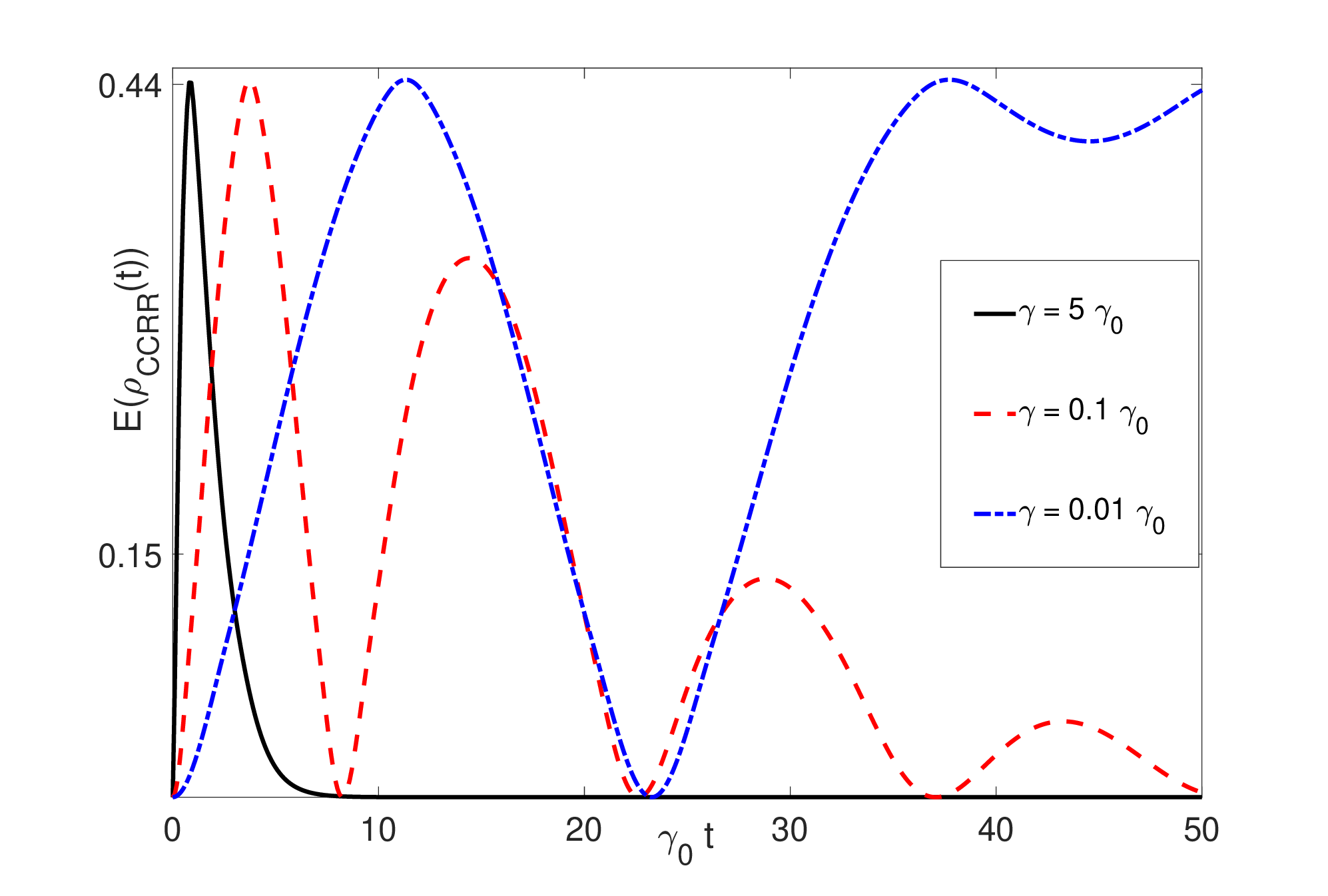}}
\caption{(Color online) Genuine entanglement for four qubits is plotted against parameter $\gamma_0 \, t$ 
with $\alpha = \sqrt{1/3}$ and $\beta = \sqrt{2/3}$. Solid (black) line is for Markovian dynamics and other
two curves are for non-Markovian dynamics.}
\label{FIG:CR3}
\end{figure}
Figure (\ref{FIG:CR3}) depicts the genuine entanglement of four qubits plotted against parameter 
$\gamma_0 \, t$ with initial condition $\alpha = \sqrt{1/3}$ and $\beta = \sqrt{2/3}$. 
For Markov process with $x = 5$ (solid curve), we can see that genuine entanglement is developed among four 
qubits immediately after $\gamma_0 \, t > 0$, reaches to its peak value and then decays to zero in 
finite time. For non-Markov process with $x = 0.1$, we see that genuine entanglement (dashed curve) exhibits 
collapse/revival phenomenon with decaying amplitudes. For $x = 0.01$ (dashed-dotted line), oscillations 
of genuine entanglement have broader duration. 

The above examples take $\beta < 2 \alpha$ with sudden birth of entanglement among reservoirs comes earlier 
than entanglement is lost from cavities. We now consider the case $\beta > 2 \alpha$ as it is known that 
sudden birth of entanglement comes later than entanglement is lost from cavities. This creates a time window 
where neither cavities nor reservoirs are entangled with each other \cite{Lopez-PRL101-2008}. In our earlier 
study, we have observed a peculiar feature of genuine entanglement for this time 
window \cite{Ali-PRA90-2014}. Here, we aim to study the effects of non-Markovian process on this sudden 
change in dynamics of genuine entanglement. 

\begin{figure}[h]
\centering
\scalebox{2.30}{\includegraphics[width=1.99in]{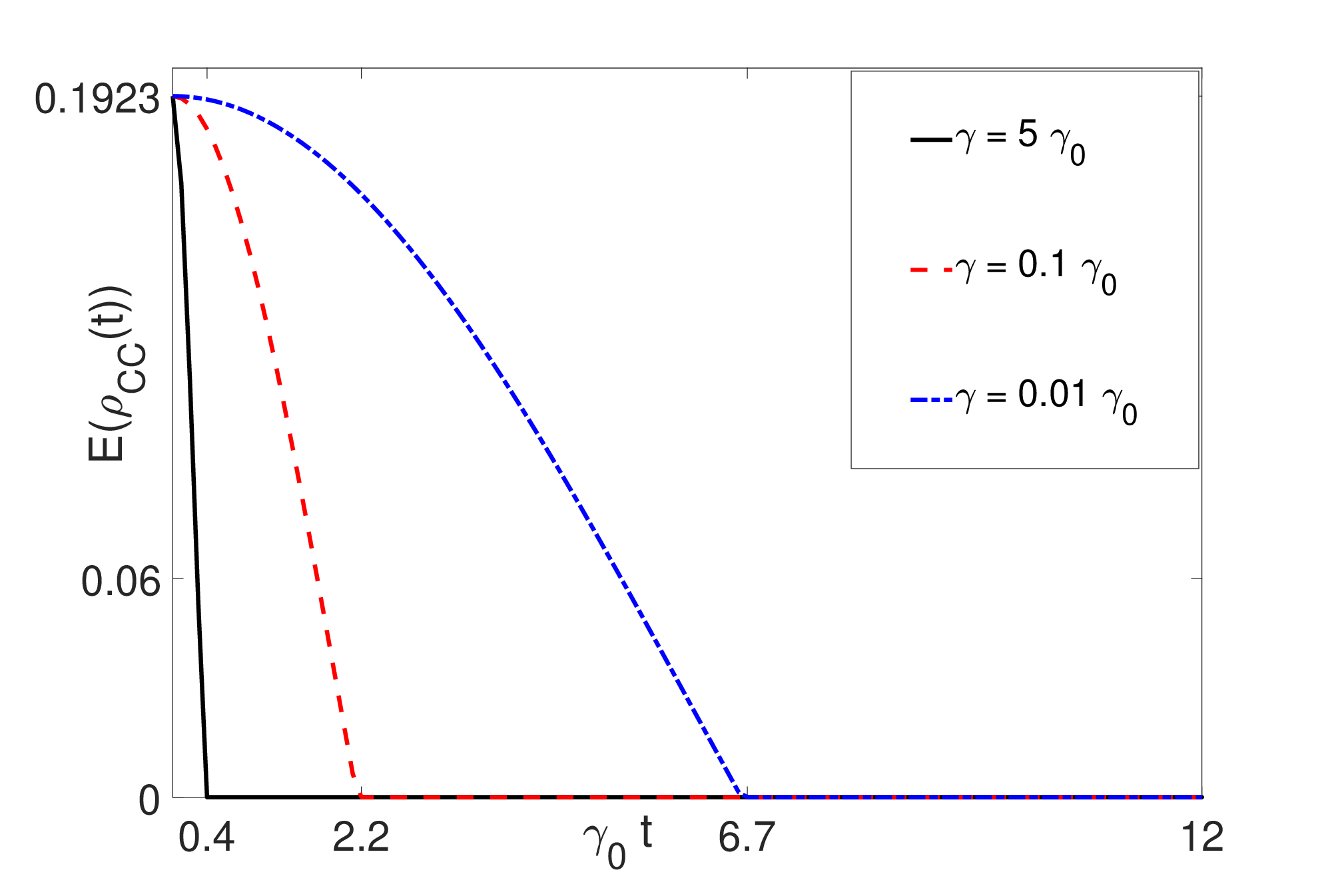}}
\caption{(Color online) Same caption as Figure (\ref{FIG:CR1}) except we take $\alpha = \sqrt{1/26}$ and 
$\beta = 5 \, \sqrt{1/26}$. See text for more descriptions.}
\label{FIG:CR4}
\end{figure}
Figure (\ref{FIG:CR4}) depicts entanglement of cavity-cavity system against parameter $\gamma_0 \, t$. We 
start with an initial state (Eq.~(\ref{Eq:JIS})) having $\alpha = \sqrt{1/26}$ and 
$\beta = 5 \, \sqrt{1/26}$. The numerical value for entanglement (negativity) for this initial state 
is $\approx 0.1923$. The entanglement of cavity qubits is lost at $\gamma_0 \, t \approx 0.4$ for 
Markov process (solid curve) with $x = 5$. For non-Markov process, other two curves indicate that 
entanglement is lost at $\gamma_0 \, t \approx 2.2$ and $\gamma_0 \, t \approx 6.7$ for $x = 0.1$ 
and $x = 0.01$, respectively. Although, Figure (\ref{FIG:CR4}) shows horizontal scale up to 
$\gamma_0 \, t = 12$, but we have checked till $\gamma_0 \, t = 50$ but there is no entanglement revival. 

\begin{figure}[h]
\scalebox{2.29}{\includegraphics[width=1.99in]{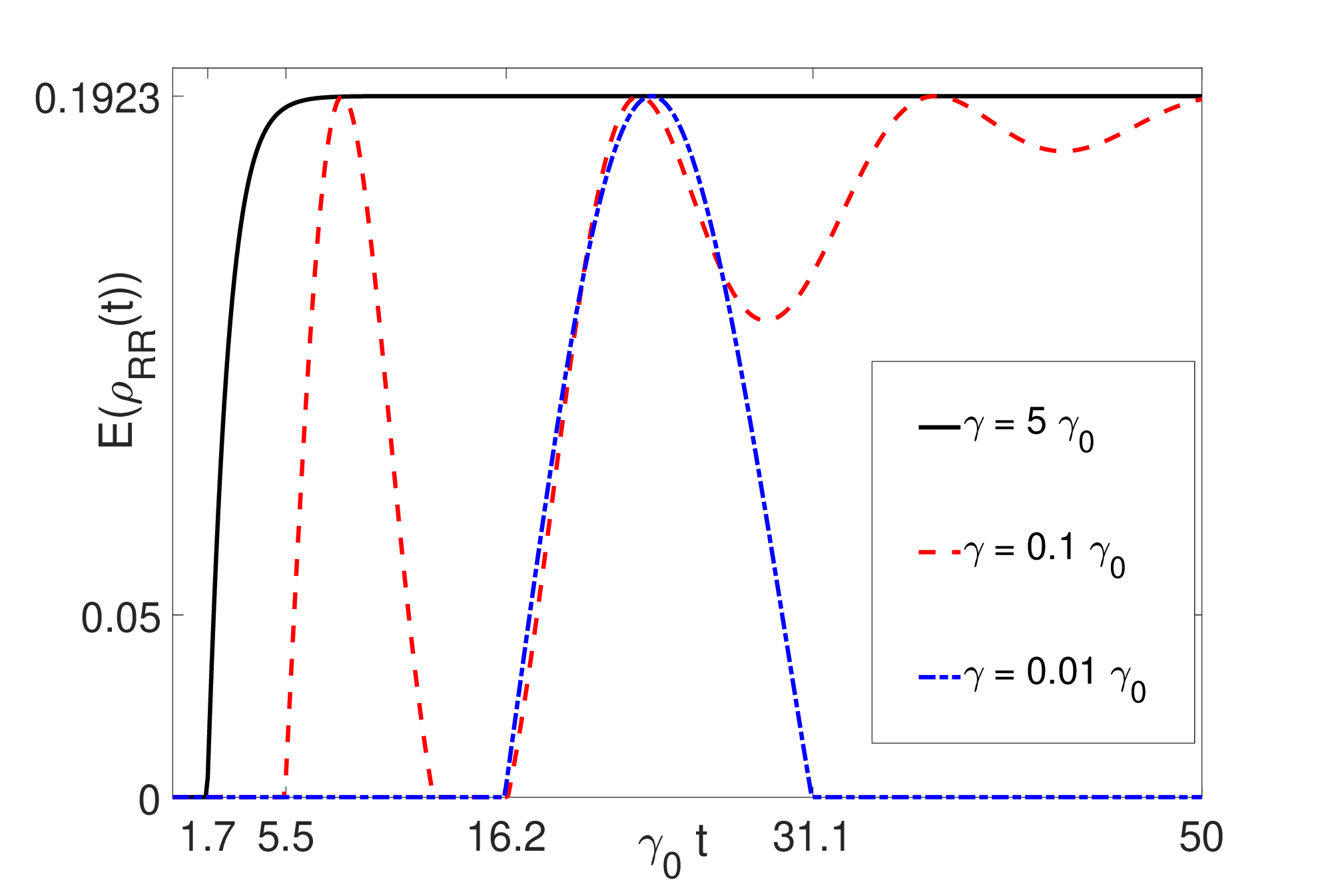}}
\caption{(Color online) Same caption as Figure (\ref{FIG:CR2}) except we take $\alpha = \sqrt{1/26}$ and 
$\beta = 5 \, \sqrt{1/26}$. See text for more descriptions.}
\label{FIG:CR5}
\end{figure}
In Figure (\ref{FIG:CR5}), we plot entanglement of reservoir-reservoir qubits against 
parameter $\gamma_0 \, t$ for an initial state (Eq.~(\ref{Eq:JIS})) having $\alpha = \sqrt{1/26}$ 
and $\beta = 5 \, \sqrt{1/26}$. We observe that in Markovian process (black solid line), entanglement 
appears at $\gamma_0 \, t \approx 1.7$ among reservoirs and reaches to its maximum value of $0.1923$ in 
finite time. We also observe that in the interval $0.4 \lesssim \gamma_0 \, t \lesssim 1.7$, neither 
cavities nor reservoirs are entangled as expected. This time window 
also occurs for non-Markovian process as it clear from Figure (\ref{FIG:CR5}). 
For $x = 0.1$, we observe that entanglement exhibits collapse and revival before 
gradually reaching its limiting value. For $x = 0.01$, entanglement appears abruptly at 
$\gamma_0 \, t \approx 16.2$ and vanishes again at $\gamma_0 \, t \approx 31.1$. We have not checked whether
it appears again after $\gamma_0 \, t = 50$.

\begin{figure}[h]
\scalebox{2.29}{\includegraphics[width=1.99in]{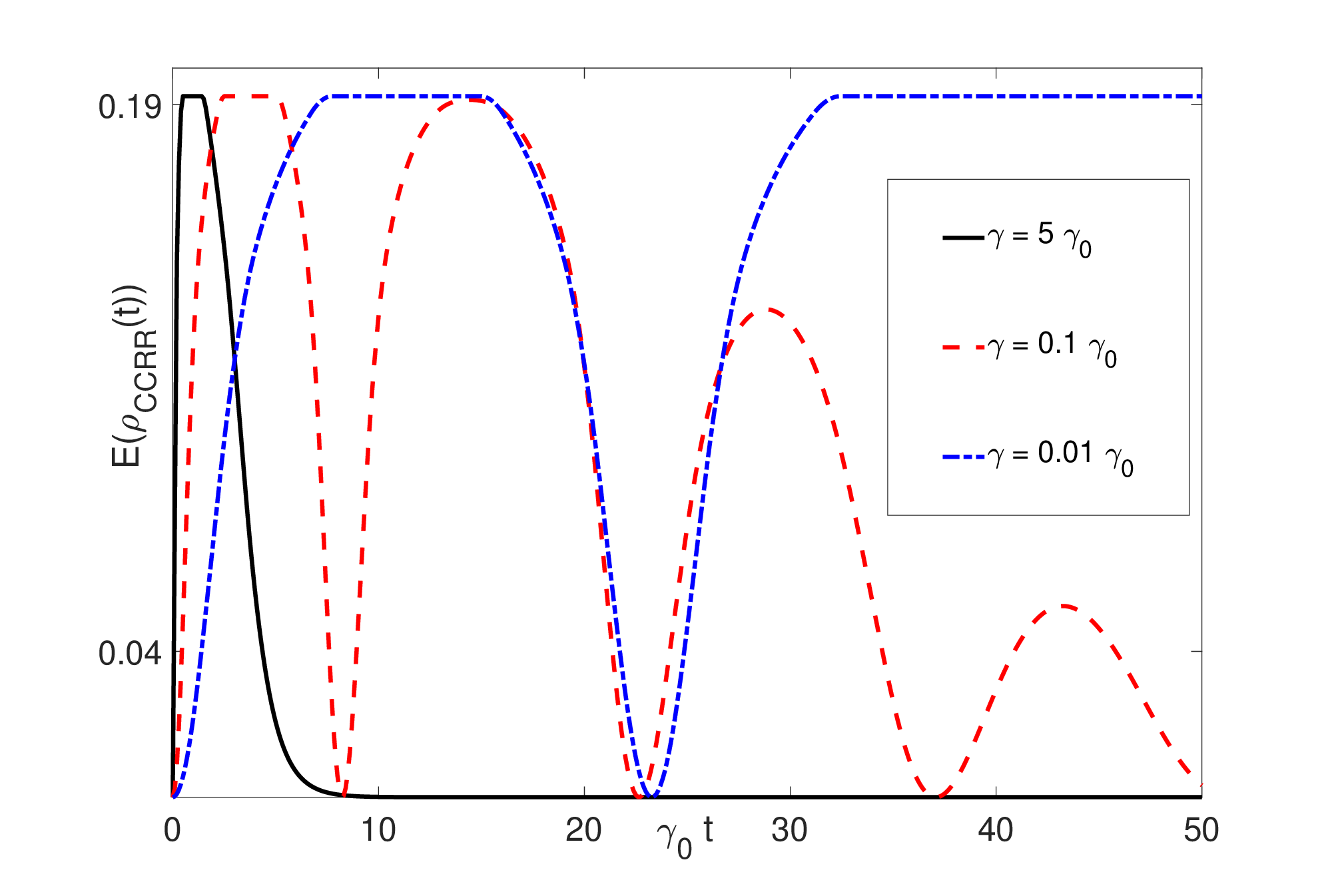}}
\caption{(Color online) Same caption as Figure (\ref{FIG:CR3}) except we take $\alpha = \sqrt{1/26}$ and 
$\beta = 5 \, \sqrt{1/26}$. See text for more descriptions.}
\label{FIG:CR6}
\end{figure}
In Figure (\ref{FIG:CR6}), we plot genuine entanglement of four qubits against parameter $\gamma_0 \, t$ 
for an initial state (Eq.~(\ref{Eq:JIS})) with $\alpha = \sqrt{1/26}$ and $\beta = 5 \, \sqrt{1/26}$. It 
is interesting to observe that sudden change in dynamics of genuine entanglement occurs not only for 
Markov process but for non-Markovian dynamics as well. We have observed sudden change 
in dynamics of genuine entanglement for Markov process in an earlier study \cite{Ali-PRA90-2014}. For 
Markov process reflected by solid curve with $x =5$, genuine entanglement starts growing among four qubits, 
reaches to a fixed value and then stops changing for some time. This interval 
$0.4 \lesssim \gamma_0 \, t \lesssim 1.5$ lies within the time window where neither cavities nor reservoirs 
are entangled. After $\gamma_0 \, t > 1.5$, genuine entanglement starts decaying to zero. 
For non-Markovian case ($x = 0.1$), such time window exists for $2.2 \lesssim \gamma_0 \, t \lesssim 5.5$ 
as evident from Figures (\ref{FIG:CR4}) and (\ref{FIG:CR5}), 
whereas genuine entanglement is locked for $2.4 \lesssim \gamma_0 \, t \lesssim 5.1$. After that, 
genuine entanglement starts decaying and exhibits collapse/revival with diminishing amplitudes. 
For dashed-dotted curve ($x = 0.01$), the time window with neither cavities nor reservoirs entangled lies 
between $6.7 \lesssim \gamma_0 \, t \lesssim 16.2$, whereas genuine entanglement exhibits sudden change 
(locked at fixed value) for $7.7 \lesssim \gamma_0 \, t \lesssim 15.2$. After this instance, genuine 
entanglement decays to zero, grows again and after $\gamma_0 \, t \gtrsim 32.4$ once again 
exhibits this sudden change (locked to same fixed value) for very long time.  

Let us take an important single parameter class of states, called Werner states, given as 
\begin{eqnarray}
\rho_p = p \, |\Phi \rangle\langle\Phi| + \frac{(1-p)}{4} \mathbb{I}_4 \, , \label{Eq:WS}
\end{eqnarray}
where $p \in [0,1]$ and $|\Phi\rangle = 1/\sqrt{2} (|0,0\rangle + |1,1\rangle)$ is the maximally entangled 
pure Bell state. Werner states are entangled for $p \in (1/3,1]$ and separable for $p \leq 1/3$. 
Under Markov process, these states always exhibit finite time disentanglement except for $p = 1$. 
Similarly, the time of appearance of entanglement among reservoir qubits is finite for $ p \in (1/3,1)$ 
and becomes zero for $p = 1$. 

\begin{figure}[h]
\scalebox{2.30}{\includegraphics[width=1.99in]{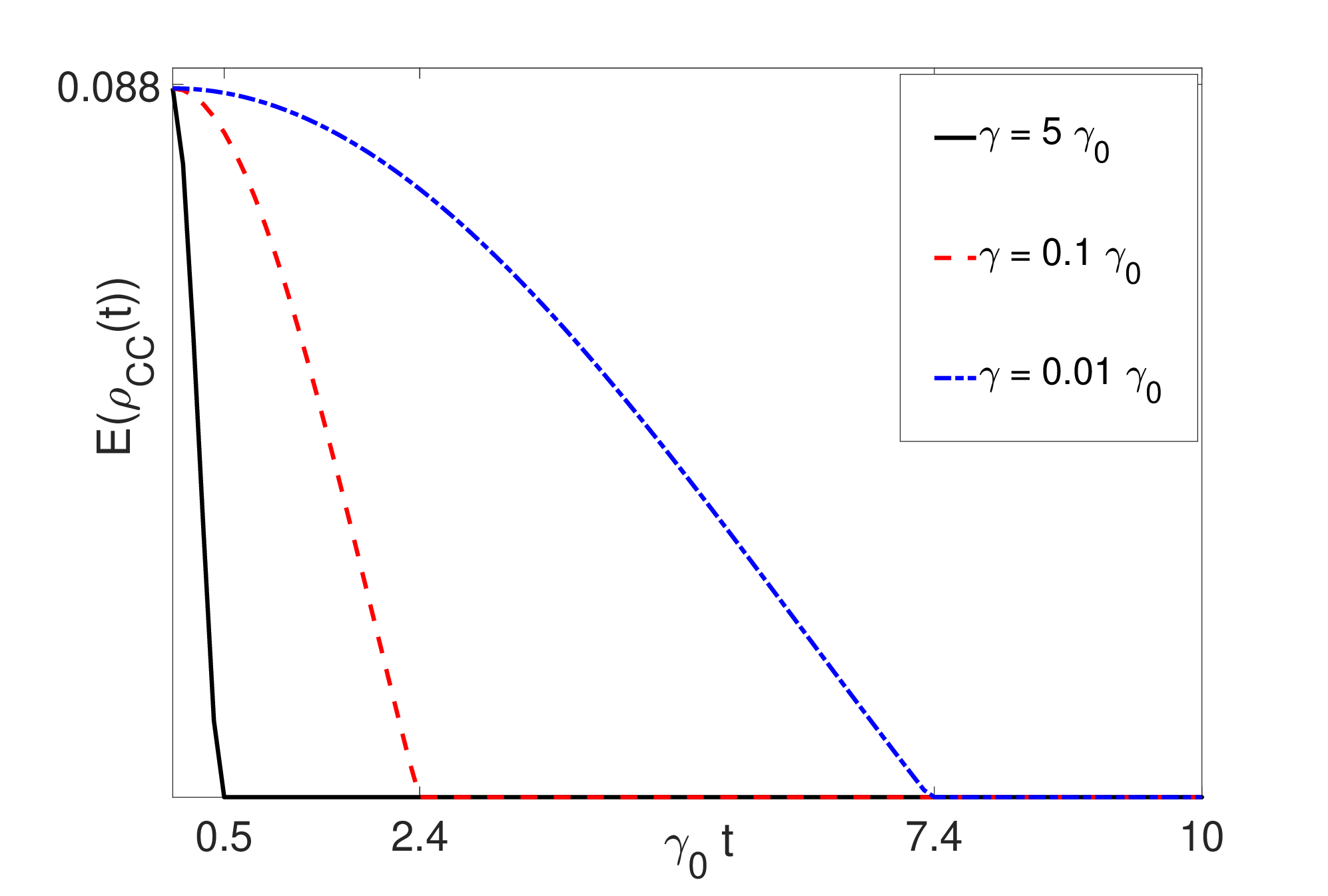}}
\caption{(Color online) Entanglement for cavity-cavity qubits is plotted against parameter $\gamma_0 \, t$ 
for Werner states with $p = 0.45$. Markov process is shown by solid curve whereas dashed line and 
dashed-dotted curves are for non-Markov process.}
\label{FIG:WSc1}
\end{figure}
In Figure (\ref{FIG:WSc1}), we plot entanglement for cavity-cavity qubits of Werner states with 
$p 0.45$ against parameter $\gamma_0 \, t$. We observe that under Markov process denoted by (black) 
solid curve, entanglement comes to an end at $\gamma_0 \, t \approx 0.5$. For non-Markov process, shown 
by (red) dashed curve and (blue) dashed-dotted 
curves, end of entanglement is delayed to $\gamma_0 \, t \approx 2.4$ and $\gamma_0 \, t \approx 7.4$, 
respectively. Although, horizontal axes is till $\gamma_0 \, t = 10$, however, we have checked that till 
$\gamma_0 \, t = 50$, there is no revival of entanglement.

\begin{figure}[h]
\scalebox{2.30}{\includegraphics[width=1.99in]{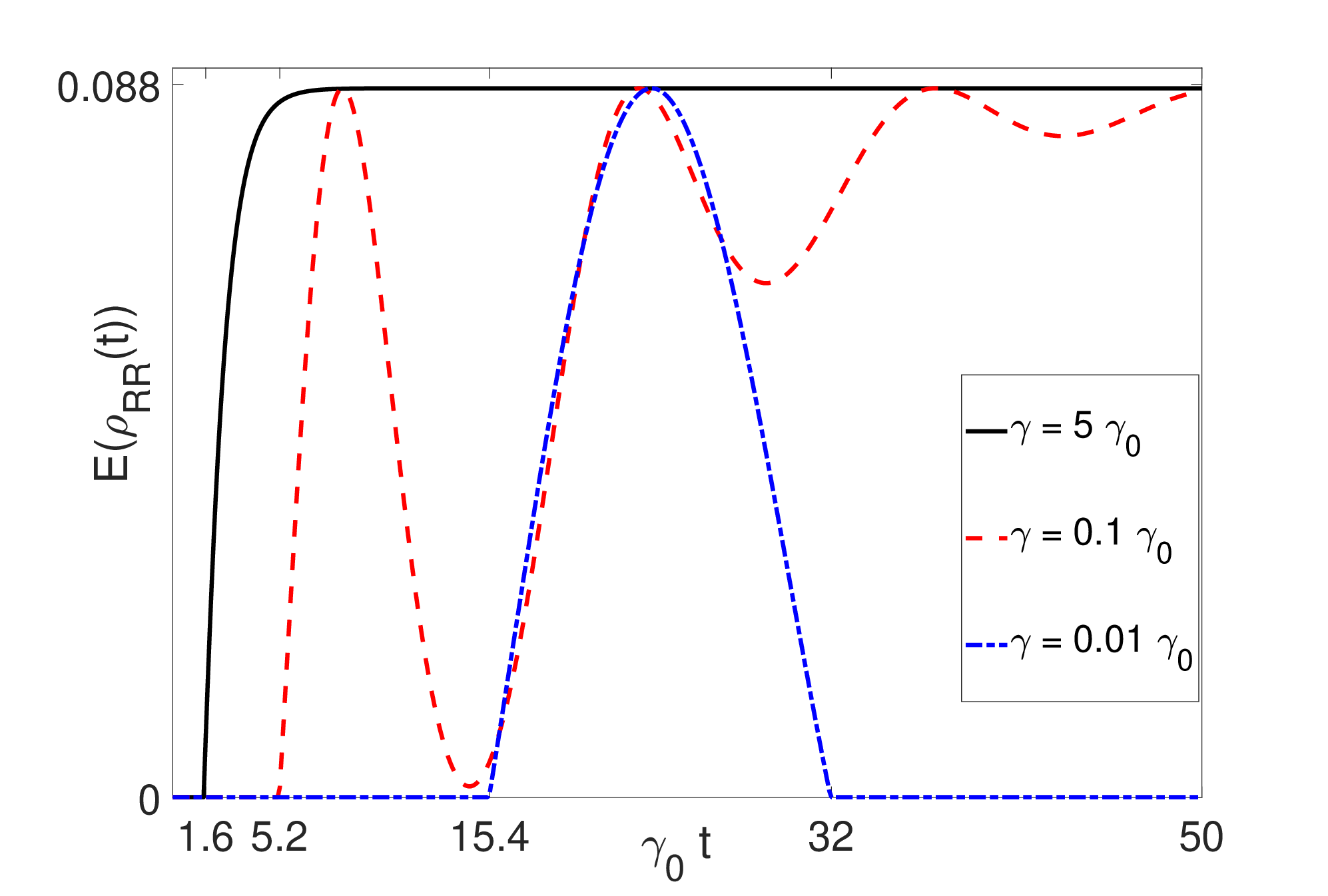}}
\caption{(Color online) Same caption as Figure (\ref{FIG:WSc1}) but for reservoir qubits.}
\label{FIG:WSc2}
\end{figure}

Figure (\ref{FIG:WSc2}) shows entanglement for reservoir-reservoir qubits for Werner states with $p = 0.45$. 
We find that entanglement appears at $\gamma_0 \, t \approx 1.6$ for Markov process, later than 
respective disappearance from cavity qubits. For non-Markov process, entanglement appears at 
$\gamma_0 \, t \approx 5.2$ (for $x = 0.1$) and $\gamma_0 \, t \approx 15.4$ (for $x = 0.01$) and have 
similar dynamics as in Figure (\ref{FIG:CR5}). All these curves are for the time windows where neither 
cavities nor reservoirs are entangled.

\begin{figure}[h]
\scalebox{2.30}{\includegraphics[width=1.99in]{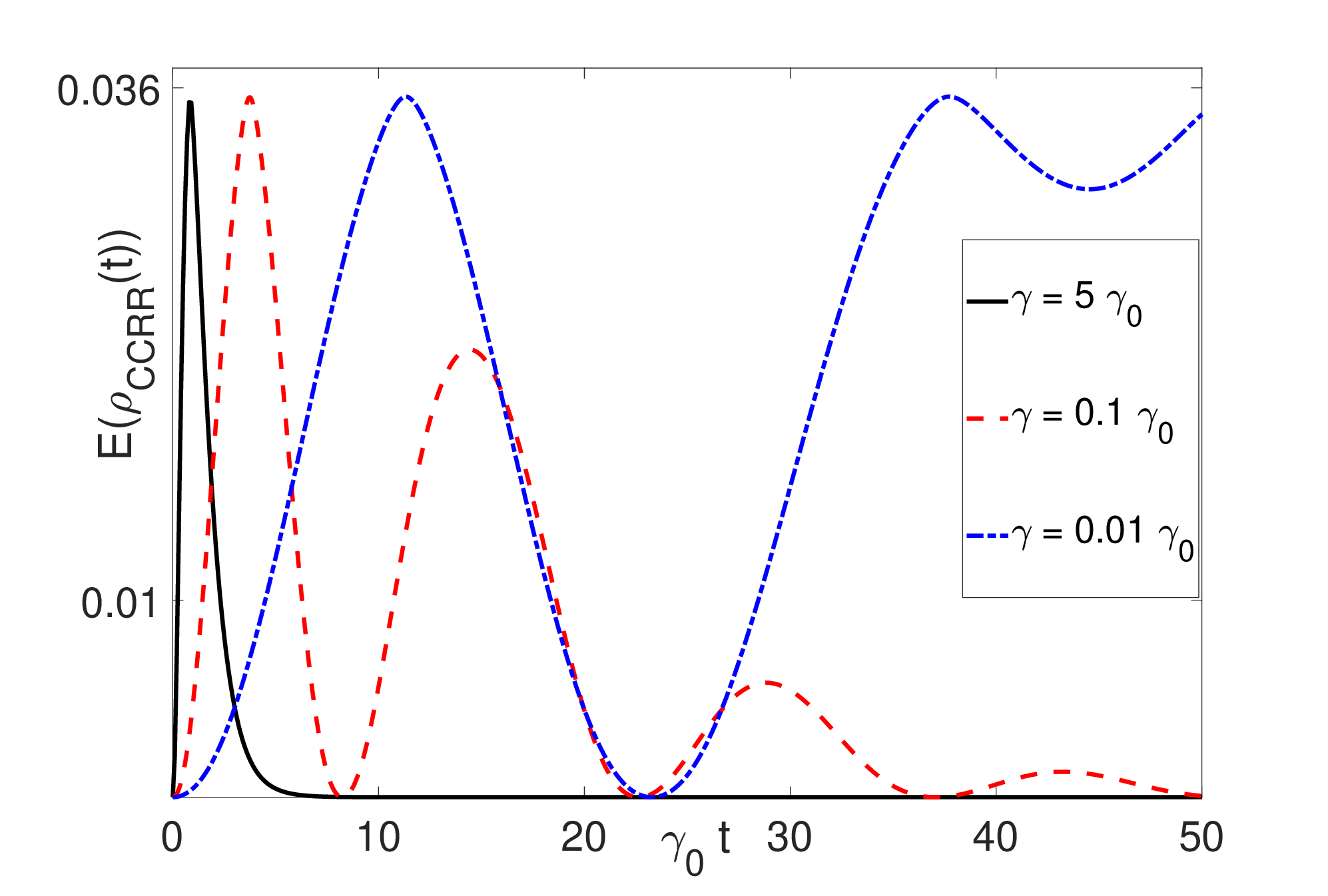}}
\caption{(Color online) Genuine entanglement is plotted against parameter $\gamma_0 \, t$ for Markov 
($x = 5$) and non-Markov ($x < 2$) process. We take Werner states with $p = 0.45$ as initial states.}
\label{FIG:WSc3}
\end{figure}
In Figure (\ref{FIG:WSc3}), we plot genuine entanglement for four qubits with initial state given by 
Eq.(\ref{Eq:WS}) with $p = 0.45$. We observe that this figure is similar to Figure (\ref{FIG:CR3}). 
Although, we have time windows with neither cavities nor reservoirs are entangled but we do not observe 
any sudden change phenomenon as we observed in Figure (\ref{FIG:CR6}). This indicates that sudden change 
in dynamics of genuine entanglement may be only related with the specific quantum states and it is not a 
generic feature. 

\begin{table}
\begin{tabular}{|  c | c | c |  c |} % centered columns (4 columns)
\hline %inserts double horizontal lines
\parbox[t]{2cm}{\textbf{Case}} & \parbox[t]{4cm}{\textbf{Cavity-Cavity qubits}} & \parbox[t]{4cm}{\textbf{Reservoir-Reservoir qubits}} 
& \parbox[t]{5cm}{\textbf{Four qubits collective state}} \\  % inserts table
\hline % inserts single horizontal line
\parbox[t]{2cm}{\textbf{Markov process}} & \parbox[t]{4cm}{Sudden death of entanglement is observed.} 
& \parbox[t]{4cm}{Sudden birth of entanglement is observed.} & \parbox[t]{5cm}{Freezing dynamics of genuine entanglement is observed (verified in experiments).} \\ % inserting body of the table
\hline
\parbox[t]{2cm}{\textbf{Non-Markov process}} & \parbox[t]{4cm}{Sudden death of entanglement is delayed along with collapse/revival phenomenon.} 
& \parbox[t]{4cm}{Sudden birth of entanglement is delayed with oscillations.} 
& \parbox[t]{5cm}{We observe recurrent freezing dynamics of genuine entanglement (yet to be studied in experiments).}  \\  
\hline %inserts single line
\end{tabular}
\caption{A comparison between markovian and non-markovian entanglement dynamics.}
\label{comp} % is used to refer this table in the text
\end{table}
In table I, we provide the comparison between entanglement dynamics in both markovian and non-markovian domain. We have compared entanglement dynamics for cavity-cavity qubits, reservoir-reservoir qubits, and four qubits collective state.

%%%%%%%%%%%%%%%%%%%%%%%%%%%%%%%% 
\section{Discussion and Summary} \label{Sec:conc}
%%%%%%%%%%%%%%%%%%%%%%%%%%%%%%%% 

We studied dynamics of bipartite and genuine multipartite entanglement under Markov and non-Markov process. 
We found that in both approximations, bipartite entanglement may vanish from cavity qubits and appear among 
reservoir qubits. Whenever there is sudden death of entanglement in cavity qubits, there is sudden birth 
of entanglement among reservoir qubits. Depending upon initial states, time of entanglement sudden birth in 
reservoirs can be before, earlier or at the same time as time of entanglement sudden death in cavities. 
Non-Markov process may bring oscillations in dynamics of bipartite entanglement as well as genuine 
entanglement. In general, we observed that non-Markov process delays the end of entanglement from cavity 
qubits.
For multipartite system of four qubits, we found that genuine entanglement develops immediately once the 
interaction starts and reaches to its peak value before decaying again. 
For a specific initial state of cavity qubits,
we observed a peculiar feature in the dynamics of genuine entanglement under both Markov and non-Markov 
process. It is remarkable that for a time window where cavity qubits as well as reservoir qubits are not 
entangled in a bipartite state, nevertheless collective state of four qubits achieves its maximum value 
of entanglement for this time window. Not only that but within this time window, genuine entanglement 
also exhibits a sudden change in its dynamics. Genuine entanglement freezes to this constant maximum value 
for some time and then starts decaying before this time window is about to end.       
Under Markov process, such sudden change happens only one time window. Under non-Markov process, sudden 
change phenomenon may occur multiple times. 
We found that sudden change behavior of genuine entanglement is observed for only one specific class of 
initial states. Other states do not exhibit such phenomenon. This may indicate that sudden change in 
dynamics of multipartite entanglement is not a generic feature. 
In addition, as we tried to analyze in our earlier study, \cite{Ali-PRA90-2014} 
with an addition of tiny amount of white noise with initial state exhibiting sudden freezing, this feature seems to wipe out. 
As the probability amplitude $\beta$ is larger than $2 \alpha$ , this indicate that some of the eigenvalues of the matrix
are locking the genuine entanglement to a fix value for some time. Unfortunately, we were unable to obtain the analytical expressions for 
the eigenvalues due to one additional parameter. However, this remains an open question to find out 
the reason of freezing dynamics of genuine entanglement related with this specific state.
We expect that our findings can be verified in experiments \cite{Aguilar-PRL113,Silva-PRA101} conducted for Markov process and beyond.
\begin{acknowledgments}
This article is inspired by the Festschrift of Prof. Dr. A. Ravi. P. Rau. I submit my profound regards to him. 
Our mutual collaboration over the years led to several articles in entanglement dynamics under decoherence. 
His ideas and insights in different areas of atomic physics and quantum information have ramifications for 
years to come. Thanks Ravi for your support and friendship. I am thankful to Dr. Chitra Rangan and Dr. Sai Vinjanampathy for inviting me to 
write an article on this occasion. This invitation triggered this entire work from start till end. I am also grateful to both referees for their constructive comments and suggested improvements. I am indeed thankful to one of the referee for educating me on 'fully inseparable states' \cite{Schneeloch-PR2020}.  
\end{acknowledgments}

\end{document}